\documentclass[12pt]{article}
\usepackage{graphics,epsfig}
\usepackage[english]{babel}
\newcommand{\cinst}[2]{$^{\mathrm{#1}}$~#2\par}
\newcommand{\crefi}[1]{$^{\mathrm{#1}}$}

\begin{document}

%\begin{titlepage}

\thispagestyle{empty}
%%%%%%%%%%%%%%%%%%%%%%%% COVER PAGE
\begingroup

%\raisebox{0.5cm}[0cm][0cm] {
%\begin{tabular*}{\hsize}{@{\hspace*{5mm}}ll@{\extracolsep{\fill}}r@{}}
%\begin{minipage}[t]{3cm}
%\vglue.5cm
%\end{minipage}
%&
%\begin{minipage}[t]{7cm}

%\end{minipage}
%&
%\begin{minipage}[t]{7cm}
%\vglue.5cm {\bf YerPhI Preprint 1614(2)-2007} %\vglue.1cm

%\end{minipage}
%\end{tabular*}
%}

\begin{center}
%{\large{YEREVAN PHYSICS INSTITUTE}}

%\hspace{9cm}{Draft 1}
\vglue 1.0cm {\Large \textbf{ Charged $\rho$ meson production in
neutrino-induced \\ \vspace{0.5cm} reactions at $\langle E_\nu
\rangle \approx$ 10 GeV}}

%(DRAFT 1)
\end{center}

\vspace{1.cm}

\begin{center}
{\large SKAT Collaboration}

 N.M.~Agababyan\crefi{1}, V.V.~Ammosov\crefi{2},
 M.~Atayan\crefi{3},\\
 N.~Grigoryan\crefi{3}, H.~Gulkanyan\crefi{3},
 A.A.~Ivanilov\crefi{2},\\ Zh.~Karamyan\crefi{3},
V.A.~Korotkov\crefi{2}

\setlength{\parskip}{0mm}
\small
%\HRule\\

\vspace{1.cm} \cinst{1}{Joint Institute for Nuclear Research,
Dubna, Russia} \cinst{2}{Institute for High Energy Physics,
Protvino, Russia} \cinst{3}{Yerevan Physics Institute, Armenia}
\end{center}
\vspace{100mm}

{\centerline{\bf YEREVAN  2008}}

%\end{titlepage}

\newpage
\vspace{1.cm}
\begin{abstract}
The neutrinoproduction of charged $\rho$ mesons on nuclei and
nucleons is investigated for the first time at moderate energies
($\langle E_\nu \rangle \approx$ 10 GeV), using the date obtained
with SKAT bubble chamber. No strong nuclear effects are observed
in $\rho^+$ and $\rho^-$ production. The fractions of charged and
neutral pions originating from $\rho$ decays are obtained and
compared with higher energy data. From analysis of the obtained
and available data on $\rho^+$ and $K^{*+}$(892)
neutrinoproduction, the strangeness suppression factor in the
quark string fragmentation is extracted: $\lambda_s =
0.18\pm0.03$. Estimations are obtained for cross sections of
quasiexclusive single $\rho^+$ and coherent $\rho^+$
neutrinoproduction on nuclei. The estimated coherent cross section
$\sigma_{\rho^+}^{coh}$ = (0.29$\pm0.16)\cdot 10^{-38}$ cm$^2$ is
compatible with theoretical predictions.

\end{abstract}

\newpage
\setcounter{page}{1}
\section{Introduction}

%\vglue2cm
In order to infer as comprehensive as possible information about
the space-time pattern of leptoproduced quark string fragmentation
processes, experimental data on nuclear targets are needed, not
only concerning the production of stable hadrons, but also
hadronic resonances. At present, more or less detailed
experimental data collected in neutrinonuclear interactions are
available only for neutral $\rho$ mesons (\cite{ref1,ref2,ref3}
and references therein), while those for charged $\rho$ mesons are
rather scarce and obtained at comparatively high energies of
(anti)neutrino, $\langle E_\nu \rangle \sim$ 40-50 GeV
\cite{ref2}. The aim of this work is to study the
neutrinoproduction of $\rho^+$ and $\rho^-$ mesons in
neutrino-nucleus ($\nu$ A) and neutrino-nucleon ($\nu$ N) charged
current interactions at intermadiate energies ($\langle
E_{\nu}\rangle \sim$ 10 GeV). In Section 2, the experimental
procedure is described. Section 3 presents the experimental data
including: a) the inclusive production of $\pi^0$ mesons; b) the
mean multiplicities of $\rho^+$ and $\rho^-$ mesons and the ratios
of $\rho$ meson to pion yields, compared to the data at higher
energies \cite{ref2}; c) the  quasiexclusive and coherent single
$\rho^+$ neutrinoproduction on nuclear targets. The results are
summarized in Section 4.

\section{Experimental procedure}

\noindent The experiment was performed with SKAT bubble chamber
\cite{ref4}, exposed to a wideband neutrino beam obtained with a
70 GeV primary protons from the Serpukhov accelerator. The chamber
was filled with a propane-freon mixture containing 87 vol\%
propane ($C_3H_8$) and 13 vol\% freon ($CF_3Br$) with the
percentage of nuclei H:C:F:Br = 67.9:26.8:4.0:1.3 \%. A 20 kG
uniform magnetic field was provided within the operating chamber
volume.
\\ Charged (CC) current interactions containing a negative muon with momentum
$p_{\mu} >$0.5 GeV/c were selected. Other negatively charged
particles were considered to be $\pi^-$ mesons. Protons with
momentum below 0.6 GeV$/c$ and a fraction of protons  with
momentum 0.6-0.85 GeV$/c$ were identified by their stopping in the
chamber. Non-identified positively charged particles were
considered to be ${\pi}^+$ mesons. Events in which errors in
measuring the momenta of all charged secondaries and photons were
less than 60\% and 100\%, respectively, were selected. The mean
relative error $\langle \Delta p/p \rangle$ in the momentum
measurement for muons, pions and gammas was, respectively, 3\%,
 6.5\% and 19\%. Each event is given a weight which corrects for
the fraction of events excluded due to improperly reconstruction.
More details concerning the experimental procedure, in particular,
the reconstruction of the neutrino energy $E_{\nu}$ can be found
in our previous publications \cite{ref5,ref6}. \\ The events with
$3< E_{\nu} <$ 30 GeV were accepted, provided that the
reconstructed mass $W$ of the hadronic system exceeds 1.8 GeV. No
restriction was imposed on the transfer momentum squared $Q^2$.
The number of accepted events was 5011 (6868 weighted events). The
mean values of the kinematical variables were $\langle E_{\nu}
\rangle$ = 9.8 GeV, $\langle W \rangle$ = 2.8 GeV, $\langle W^2
\rangle$ = 8.7 GeV$^2$,
$\langle Q^2 \rangle$ = 2.6 (GeV/c)$^2$. \\
Further, the whole event sample was subdivided, using several
topological and kinematical criteria \cite{ref6,ref7}, into three
subsamples: the 'cascade' subsample $B_S$ with a sign of
intranuclear secondary interaction, the 'quasiproton' ($B_p$) and
'quasineutron' ($B_n$) subsamples. About 40\% of subsample $B_p$
is contributed by interactions with free hydrogen. Weighting the
'quasiproton' events with a factor of 0.6, one can compose a
'pure' nuclear subsample $B_A = B_S + B_n + 0.6 B_p$ and a
'quasinucleon' subsample $B_N = B_n + 0.6 B_p$. It has been
verified \cite{ref7,ref8}, that the multiplicity and spectral
characteristics of secondary particles in the $B_p (B_N)$
subsample are in satisfactory agreement with those measured with a
pure proton (deuteron) target. The effective atomic weight
corresponding to the subsample $B_A$ is estimated \cite{ref6} to
be approximately equal to $A_{eff} = 21 \pm 2$, when taking into
account the probability of secondary intranuclear interactions in
the composite target.

\section{Results}

a) The inclusive production of $\pi^0$ meson

The proper reconstruction of $\pi^0 \rightarrow 2\gamma$ decays is
the main prerequisite for that of $\rho^{\pm} \rightarrow \pi^0
\pi^{\pm}$ decays. Fig. 1 shows the two-gamma effective mass
distribution for the total sample of events corrected for losses
of non-registered or rejected $\gamma$-quanta (see \cite{ref10}
for details). The $m_{\gamma\gamma}$ - distribution was fitted as
a sum of non-relativistic Breit-Wigner function, with the pole
mass $m_0$ and width $\Gamma_0$ as free parameters, and the
background distribution of the form
\begin{equation}
BG_0 \sim m_{\gamma\gamma}^{\alpha} \, \exp \, (\beta
m_{\gamma\gamma} + \gamma m_{\gamma\gamma}^2) \, \, ,
\end{equation}
\noindent with $\alpha$, $\beta$ and $\gamma$ as free parameters.
The fitted values of $m_0 = 0.139 \pm 0.002$ GeV$/c^2$ and
$\Gamma_0 = 0.055 \pm 0.007$ GeV turn out to be compatible with
the $\pi^0$ mass and experimental resolution (estimated from
Monte-Carlo simulation), respectively. A similar fitting
procedure, but with fixed $m_0$ = 0.135 GeV$/c^2$ and $\Gamma_0$ =
0.055 GeV, was applied for the quasinucleon and nuclear
subsamples, resulting in the $\pi^0$ mean multiplicities $\langle
n_{\pi^0} \rangle_N = 0.787 \pm 0.080$ and $\langle n_{\pi^0}
\rangle_A = 0.904 \pm 0.066$, respectively. The nuclear
enhancement factor for the $\pi^0$ yield, $R_{\pi^0} = \langle
n_{\pi^0}\rangle_A/\langle n_{\pi^0}\rangle_N = 1.15 \pm 0.14$,
practically coincides with that for $\pi^-$ meson, $R_{\pi^-} =
1.12 \pm 0.03$ (see also \cite{ref3, ref8, ref9}), quantitatively
explained by
secondary intranuclear interaction processes \cite{ref8,ref9}. \\
In the next subsection, where the effective mass distributions of
$\pi^+{\gamma\gamma}$ and $\pi^-{\gamma\gamma}$ systems are
analyzed, only two-gamma combinations with 0.105 $<
m_{\gamma\gamma} <$ 0.165 GeV$/c^2$ will be kept.

b) Mean multiplicities of charged $\rho$ mesons

\noindent Fig. 2 shows the effective mass distributions for
$\pi^+{\gamma\gamma}$ and $\pi^-{\gamma\gamma}$ systems, both for
nuclear and quasinucleon subsamples, corrected for losses of
reconstructed $\pi^0$ and contamination from the background
$\gamma\gamma$ combinations. A clear signal for $\rho^+$, but a
faint one for $\rho^-$ are seen. The distribution were fitted by
the form
\begin{equation}
dN/dm = BG \cdot \, (1 + \, \alpha_{\rho} \, BW_{\rho}) \, \, ,
\end{equation}
\noindent where the mass dependence of the background was
parametrized as
\begin{equation}
BG = {\alpha}_1 \exp \, ({\alpha}_2 m + {\alpha}_3 m^2) \, \, ,
\end{equation}
\noindent with $m \equiv m_{\pi^+ {\gamma\gamma}}$\, or\,
$m_{\pi^- {\gamma\gamma}}$ and $\alpha_1$, $\alpha_2$, $\alpha_3$
being free parameters, while for $BW_{\rho}$ the relativistic
Breit-Wigner function \cite{ref11} was used, with fixed pole mass
$m_{\rho}$ = 0.776 GeV$/c^2$ and width $\Gamma_{\rho}$ = 0.149 GeV
\cite{ref12}. The experimental mass resolution was taken into
account by replacing $\Gamma_{\rho}$ by $\sqrt{\Gamma_{\rho}^2 +
\Gamma_{res}^2}$ with $\Gamma_{res}$ = 0.11 GeV estimated from
Monte-Carlo simulations.
\\ A similar analysis was also performed for the $\pi^+
{\gamma\gamma}$ system produced in the forward ($x_F > 0, x_F$
being the Feynman variable) and backward ($x_F < 0$) hemispheres
in the hadronic c.m.s. \\ The resulting mean yields of charged
$\rho$ mesons are presented in Table 1. It should be emphasized,
that the sum of extracted yield of $\rho^+$ at $x_F < 0$ and $x_F
> 0$ is in good agreement with the total yield. As expected, the
yield of unfavorable $\rho^-$ (not containing the current quark)
is much smaller than that of favorable $\rho^+$ (which can contain
the current quark). The data do not reveal, due to comparatively
large statistical errors, significant nuclear effects in charged
$\rho$ neutrinoproduction, perhaps except a faint indication on
the shifting the $\rho^+$ $x_F$ - distribution in the nuclear
interactions towards backward hemisphere as compared to
quasinucleon interactions. A similar effect, but better expressed,
was observed recently for $\rho^0$ mesons, indicating on a
non-negligible role of secondary intranuclear interaction
processes \cite{ref3}.

\begin{table}[ht]
\caption{The mean multiplicities of $\rho^+$ and $\rho^-$ mesons
in quasinucleon and nuclear interactions.}
\begin{center}
\begin{tabular}{|l|c|c|c|c|}
  \hline
\multicolumn{4}{|c|}{$\langle n_{\rho^+}\rangle$}
& \multicolumn{1}{c|}{$\langle n_{\rho^-}\rangle$} \\
\hline
% &\multicolumn{4}{c|}{4 $< W^2 <$ 25
%GeV$^2$}\\  \multicolumn{5}{|c|}{} \\ $h^+(x_F > 0)$
$A$&all $x_F$&$x_F < 0$&$x_F >0$& all $x_F$
\\ \hline
1&0.101$\pm$0.032&0.046$\pm$0.033&0.061$\pm$0.030 &0.026$\pm$0.024
\\
21&0.120$\pm$0.031&0.067$\pm$0.025&0.051$\pm$0.015
&0.039$\pm$0.015
\\
 \hline

\end{tabular}
\end{center}
\end{table}

c) The $W$- dependence of the $\rho$ yield and the ratio of $\rho$
and pion yields

\noindent Fig. 3 shows the $W$ - dependence of the charged $\rho$
yields measured in this work for $\nu$A - interactions at $W$= 2.8
GeV and in \cite{ref2} for $\nu (\bar{\nu}) Ne$ - interactions.
For comparison, the data for favorable vector mesons $\rho^0$
\cite{ref2,ref3} and $K^{*+}$(892) (\cite{ref13} and references
therein) are also plotted. The $W$ - dependence of the mean yields
can be approximately described by a simplest linear form $b \cdot
(W-W_0)$ at the fixed threshold value $W_0$ = 1.8 GeV. The fitted
slope parameters $b$ are given in Table 2.

\begin{table}[ht]
\caption{The slope parameter $b$ (in GeV$^{-1}$).}
\begin{center}
\begin{tabular}{|l|c|c|c|}
  \hline
%\multicolumn{5}{|c|}{} \\
% &\multicolumn{4}{c|}{4 $< W^2 <$ 25
%GeV$^2$}\\  \multicolumn{5}{|c|}{} \\ $h^+(x_F > 0)$
Reaction&all $x_F$&$x_F < 0$ &$x_F > 0$
\\ \hline
 $\nu \rightarrow \rho^+$ or $\bar{\nu} \rightarrow \rho^-$
 &0.091$\pm$0.011&0.037$\pm$0.007&0.046$\pm$0.006 \\
\hline
$\nu \rightarrow \rho^0$ or $\bar{\nu} \rightarrow \rho^0$
 &0.056$\pm$0.007&0.014$\pm$0.005&0.045$\pm$0.004 \\
 \hline
\multicolumn{1}{|c|}{$\nu \rightarrow \rho^-$ or $\bar{\nu}
\rightarrow
\rho^+$} & \multicolumn{1}{c|}{0.021$\pm$0.007} & \multicolumn{2}{c|}{}\\
\hline
\multicolumn{1}{|c|}{$\nu \rightarrow K^{*+}$(892)}
& \multicolumn{1}{c|}{0.017$\pm$0.002} & \multicolumn{2}{c|}{}\\
\hline

\end{tabular}
\end{center}
\end{table}

\noindent As it is seen from Table 2 and Fig. 3, the yields of
favorable charged and neutral $\rho$'s are the same in the forward
hemisphere, testifying that the current quark $u$ (or $\bar{d}$)
recombines with sea antiquarks (quarks) $\bar{u}$ and $\bar{d}$
(or $d$ and $u$) with equal probabilities. On the other hand, the
yield of the favorable charged $\rho$ exceeds that of $\rho^0$ in
the backward hemisphere. \\ The ratio $b (K^{*+})/b (\rho^+)$ can
serve as an almost direct estimation for the strange quark
suppression factor $\lambda_s \equiv (\bar{s}/\bar{u})$ during the
string fragmentation at the considered $W$- range: $\lambda_s =
0.18 \pm 0.03$. This value is compatible with estimations
extracted from hadronic interactions (\cite{ref14,ref15} and
references therein). \\
\noindent Fig. 4 presents the $\langle n_{\rho}\rangle_A/\langle
n_{\pi}\rangle_A$ ratios extracted from \cite{ref2} and obtained
in this work, using the measured pion yields $\langle
n_{\pi^0}\rangle_A = 0.904\pm0.066$, $\langle n_{\pi^-}\rangle_A =
0.652\pm0.010$ and $\langle n_{\pi^+}\rangle_A = 1.55\pm0.06$ (the
latter error being caused mainly by the uncertainty in the
extraction of the contamination from non-identified protons). As
it is seen from the left panel of Fig. 4, the fraction of $\pi^0$
originating from decays of favorable and unfavorable $\rho$'s is
practically independent of $\langle W \rangle$ (in the considered
range $2.8 < \langle W \rangle < 4.8$ GeV) and composes on an
average $14.2\pm1.8$ and $3.4\pm1.0$\%, respectively (indicated by
dashed lines in Fig. 4). The latter value is close to the fraction
of charged pions from the decay of unfavorable $\rho$, composing
on an average $4.3\pm1.3$\% (open symbols and the dashed line in
the right panel of Fig. 4). Unlike the latter and neutral pions,
the fraction of charged pions originating from the favorable
$\rho$ decay tends to increase with $\langle W \rangle$, varying
from $7.7\pm2.0$\% at $\langle W \rangle$ = 2.8 GeV to
$12.6\pm2.9$\% at $\langle W \rangle$ = 4.8 GeV (closed symbols in
the right panel of Fig. 4).

d) Quasiexclusive and coherent single $\rho^+$ production on
nuclei

This subsection is devoted to the estimation of cross sections of
quasiexclusive and coherent single $\rho^+$ production in the
reaction
\begin{equation}
\nu A \rightarrow {\mu^-} + ({\pi^+} {\gamma\gamma}) + X_0 \, \, ,
\end{equation}
\noindent where the state $X_0$ does not contain any visible
tracks. \\
\noindent The rate of the quasiexclusive reaction
\begin{equation}
\nu A \rightarrow {\mu^-} + ({\pi^+} {\pi^0)} + X_0
\end{equation}
\noindent extracted from the $m_{\gamma\gamma}$ - distribution,
plotted in Fig. 5a, is equal to $R(\nu A \rightarrow
{\mu^-}{\pi^+}{\pi^0}X_0) = (2.4\pm0.4) \cdot 10^{-2}$. The
contamination to the latter from reactions where one or more
additional (not registered) $\pi^0$'s are produced was estimated
from the analysis of events containing more than two detected
gammas. This contamination to the reaction (5) turns out to be
rather small, (11$\pm$10)\%. As it is seen from Fig. 5a, the
$\pi^0$ mass region is practically free of background. This
observation allows one to apply less severe cut on the 'usefull'
range of $m_{\gamma\gamma}$
 , 0.053 $< m_{\gamma\gamma} <$ 0.217 GeV$/c^2$, thus reducing
the $\pi^0$ losses in the ${\pi^+} {\gamma\gamma}$ effective mass
distribution. \\
The rate $R({\mu^-}{\rho^+} X_0) = \sigma (\nu A \rightarrow
{\mu^-}{\rho^+} X_0)/\sigma (\nu A \rightarrow {\mu^-} X)$ of the
quasiexclusive incoherent reaction
\begin{equation}
\nu A \rightarrow {\mu^-}{\rho^+} X_0
\end{equation}
\noindent was estimated from the $\pi^+ {\gamma\gamma}$ effective
mass distribution (Fig. 5b) from which the events-candidates to
the $\rho^+$ coherent production (see below) were excluded. This
distribution was corrected and fitted as described above in
Subsection 3b resulting in $R({\mu^-}{\rho^+} X_0) = (1.0\pm0.5)
\cdot 10^{-2}$. The latter composes only (8.5$\pm$4.3)\% of the
$\rho^+$ production total rate (cf. Table 1) and corresponds to
the cross section $\sigma (\nu A \rightarrow {\mu^-}{\rho^+} X_0)
= (2.2\pm1.2)\cdot 10^{-38}$ cm$^2$ obtained at calculated $\nu A$
cross section $\sigma (\nu A \rightarrow {\mu^-} X) =
(1.16\pm0.09)\cdot10^{-36}$ cm$^2$ at $\langle E_{\nu} \rangle$ =
9.8 GeV, using the known neutrino-nucleon CC cross sections
$\sigma_{{\nu} p}^{CC}/E_{\nu} = (0.50\pm0.04)\cdot 10^{-38}$
cm$^2$/GeV and $\sigma_{{\nu}n}^{CC} = 2\sigma_{{\nu}p}^{CC}$
(\cite{ref12,ref16} and references therein) and taking into
account the nuclei content of the composite target. \\ As the
process (6) occurs on intranuclear neutrons, the number of events
belonging to that reaction can be normalized to the number of
events of quasineutron subsample (see Section 3), leading to an
estimation of the rate $R({\mu^-}{\rho^+} n) = \sigma (\nu n
\rightarrow {\mu^-}{\rho^+} n)/\sigma (\nu n \rightarrow {\mu^-}
X) = (4.4\pm2.2) \cdot 10^{-2}$. Note, that this value (estimated
for the first time for $\nu n$- interactions) turns out to be
significantly larger than for reaction $\nu p\rightarrow \mu^-
\rho^+ p$ on free protons, measured at a higher energy range
($\langle E_\nu \rangle$ = 15 GeV), $R({\mu^+}{\rho^+} p) =
(0.7\pm0.2) \cdot 10^{-2}$ \cite{ref17}.
\\ A particular case of quasiexclusive reaction (4) is
the exclusive reaction of coherent $\rho^+$ neutrinoproduction
\begin{equation}
\nu A \rightarrow {\mu^-}{\rho^+} A \, \, ,
\end{equation}
\noindent where the final nucleus remains in its ground state.
This process is characterized by very small values of the squared
four-momentum transfer $|\,t| < t_{max} \approx 1/R_A^2$ to the
nucleus of radius $R_A$. Even for the case of the lightest target
nucleus ($^{12}C$), $t_{max}$ = 0.006 (GeV$/c)^2$ being much
smaller than the experimental resolution estimated from
Monte-Carlo simulations, $\sigma(t) \approx$ 0.06 (GeV$/c)^2$. \\
The squared four-momentum transfer $|\,t|$ was estimated as
\begin{equation}
|\,t| = {[\sum_{i}(E_i - p_i^L)]}^2 + {(\sum_{i}\textbf{p}^t _i \,
\,)}^2 \, \, ,
\end{equation}
\noindent where $E_i$, $p_i^L$ and $\textbf{p}_i^t$ are,
respectively, the energy, longitudinal and transverse momenta of
detected final particles in the reaction (4). The $|\,t|$ -
distribution is shown in Fig. 5c, together with that for
background (incoherent) events of reaction (4) containing
observable products of the target nucleus desintegration not taken
into account in evaluation of $|\,t|$. The background distribution
is normalized to the former one at $|\,t| >$ 0.09 (GeV$/c)^2$. As
it is seen, coherent-like and incoherent events are distributed
similarly at $|\,t| >$ 0.09(GeV$/c)^2$, while in the 'coherence'
region $|\,t| < $ 0.09 (GeV$/c)^2$ there is a clear excess of
coherent-like events (8 events) over incoherent one (one event).
It has been also verified, via analysis of events with more than
two detected $\gamma$'s, that the sample of coherent-like events
is not
contaminated by processes of two or more $\pi^0$ production. \\
For all selected 8 events, the effective mass
$m_{\pi^+{\gamma\gamma}}$ turns out to be enclosed in the region
near the $\rho$ pole mass, 0.48 $\leq m_{\pi^+{\gamma\gamma}}
\leq$ 1.08 GeV$/c^2$, with the mean value $\langle
m_{\pi^+{\gamma\gamma}} \rangle$ =0.7 GeV$/c^2$ (Fig. 5d). This
allows to assume, that the most part of these events originate
from coherent reaction. The upper (lower) limit of the rate
$R_{\rho^+}^{coh}$ can be estimated assuming that all 8 events (6
events with 0.58 $\leq m_{\pi^+{\gamma\gamma}} \leq$ 1.08
GeV$/c^2$) correspond to coherent $\rho^+$ production. Excluding
the contamination from one incoherent event, as well as
introducing a correction for losses due to the cut $|\,t|
<t_{max}$ (composing 8\%), one obtains for upper and lower limits:
$R_{\rho^+}^{coh}(up) = (0.28\pm0.12)\cdot 10^{-2}$, and
$R_{\rho^+}^{coh}(low) = (0.20\pm0.11)\cdot 10^{-2}$. Note, that
the average value of these estimations, $R_{\rho^+}^{coh}(mean) =
(0.24\pm0.12)\cdot 10^{-2}$ coincides practically with the value
$R_{\rho^+}^{coh}(fit) = (0.26\pm0.14)\cdot 10^{-2}$ extracted
from the fit of data plotted in Fig. 5d by the relativistic
Breit-Wigner distribution, taking also into account the
contamination from one background event and the losses at $|\,t|
> t_{max}$. As the final estimation for $R_{\rho^+}^{coh}$, we
take the medial value $R_{\rho^+}^{coh} = (0.25\pm0.14)\cdot
10^{-2}$. Note, that the latter composes, respectively,
(2.1$\pm$1.3)\% and (4.9$\pm$3.1)\% of the total and forward
$\rho^+$ production rates (cf. Table 1). \\ The estimated rate
$R_{\rho^+}^{coh} = (0.25\pm0.14)\cdot 10^{-2}$ can be compared
with that obtained at higher-energy $\nu Ne$ interactions
\cite{ref16}, $R_{\rho^+}^{coh}(10 < E_\nu < 320$ GeV) =
($0.28\pm0.10)\cdot 10^{-2}$. The quoted values do not contradict
the expected slow variation of this rate ($0.15 \cdot 10^{-2} <
R_{\rho+}^{coh} < 0.31\cdot 10^{-2}$) in the wide range of $10 <
E_{\nu} < 300$ GeV, predicted on the basis of the vector dominance
model (suggesting $\rho$ meson dominance in the weak vector
current) and Glauber theory (\cite{ref18} and references therein).
\\ The coherent $\rho^+$ production cross section per target
nucleus is equal to $\sigma_{\rho^+}^{coh}=$(0.29$\pm$0.16)$\cdot
10^{-38}$ cm $^2$. This value, plotted in Fig. 6 together with
available data at higher $E_{\nu}$ \cite{ref19,ref20}, is
compatible with the results of theoretical calculations
\cite{ref19} done in the framework of the $\rho$ - dominance model
for the case of $Ne$ nucleus (the curve in Fig. 6).

\section{Summary}

The charged $\rho$ meson neutrinoproduction on nuclei and nucleons
is investigated for the first time at moderate energies ($\langle
E_\nu \rangle \approx$ 10 GeV). No strong nuclear effects are
observed in $\rho^+$ and $\rho^-$ production. \\ The fractions of
$\pi^0$ mesons originating from the decay of favorable $(\rho^+)$
and that of $\pi^0$ and $\pi^-$ mesons from the decay of
unfavorable $(\rho^-)$ $\rho$ mesons are found to be,
respectively, 13.3$\pm$3.6, 4.3$\pm$1.7 and 5.9$\pm$2.3\%, quite
compatible with those obtained at higher energies, while this
fraction for charged pions from favorable charged $\rho$ tends to
increase with $W$, varying from 7.7$\pm$2.0\% at $\langle W
\rangle$ = 2.8 GeV
(this work) to 12.6$\pm$2.9\% at $\langle W \rangle$ = 4.8 GeV \cite{ref2}. \\
From the obtained and available data on $\nu(\bar{\nu})A
\rightarrow \rho X$ reactions at $\langle W \rangle$ = 2.8$-$4.8
GeV, an indication is obtained that the yields of charged and
neutral favorable $\rho$ mesons in the forward hemisphere are
practically the same, thus verifying that the struct current quark
(antiquark) recombines with up and down sea antiquarks(quarks)
with equal probabilities. \\ From analysis of the available data
on $\rho^+$ and $K^{*+}$(892) neutrinoproduction at $\langle W
\rangle$ = 2.8$-$5.5 GeV, an estimation is extracted for the
strangeness suppression factor in the quark
string fragmentation process: $\lambda_s = 0.18\pm0.03$. \\
For the first time, the rate of the single $\rho^+$ production
reaction on the neutron is estimated, $R({\mu^-}{\rho^+} n) =
(4.4\pm2.2) \cdot 10^{-2}$. \\
An estimation is inferred for the cross section (per nucleus of
the composite target) of the coherent $\rho^+$ neutrinoproduction
on nuclei, $\sigma_{\rho^+}^{coh}$ = (0.29$\pm0.16)\cdot 10^{-38}$
cm$^2$. The estimated coherent cross section is compatible with
theoretical predictions based on the vector
dominance model and the Glauber theory. \\
{\bf Acknowledgement.} The activity of one of the autors (H.G.) is
supported by Cooperation Agreement between DESY and YerPhI signed
on December 6, 2002. The autors from YerPhI acknowledge the
supporting grants of Calouste Gulbenkian Foundation and Swiss
Fonds "Kidagan".

%%%%%%%%%%%%%%%%%%%%%%%%%%%%%%%%%%%%%%%%%%%%%%%%%%%%%%%%%%
   %%% References
%%%%%%%%%%%%%%%%%%%%%%%%%%%%%%%%%%%%%%%%%%%%%%%%%%%%%%%%%%

\newpage
\begin{figure}[ht]
 \resizebox{0.9\textwidth}{!}{\includegraphics*[bb =20 160 600
610]{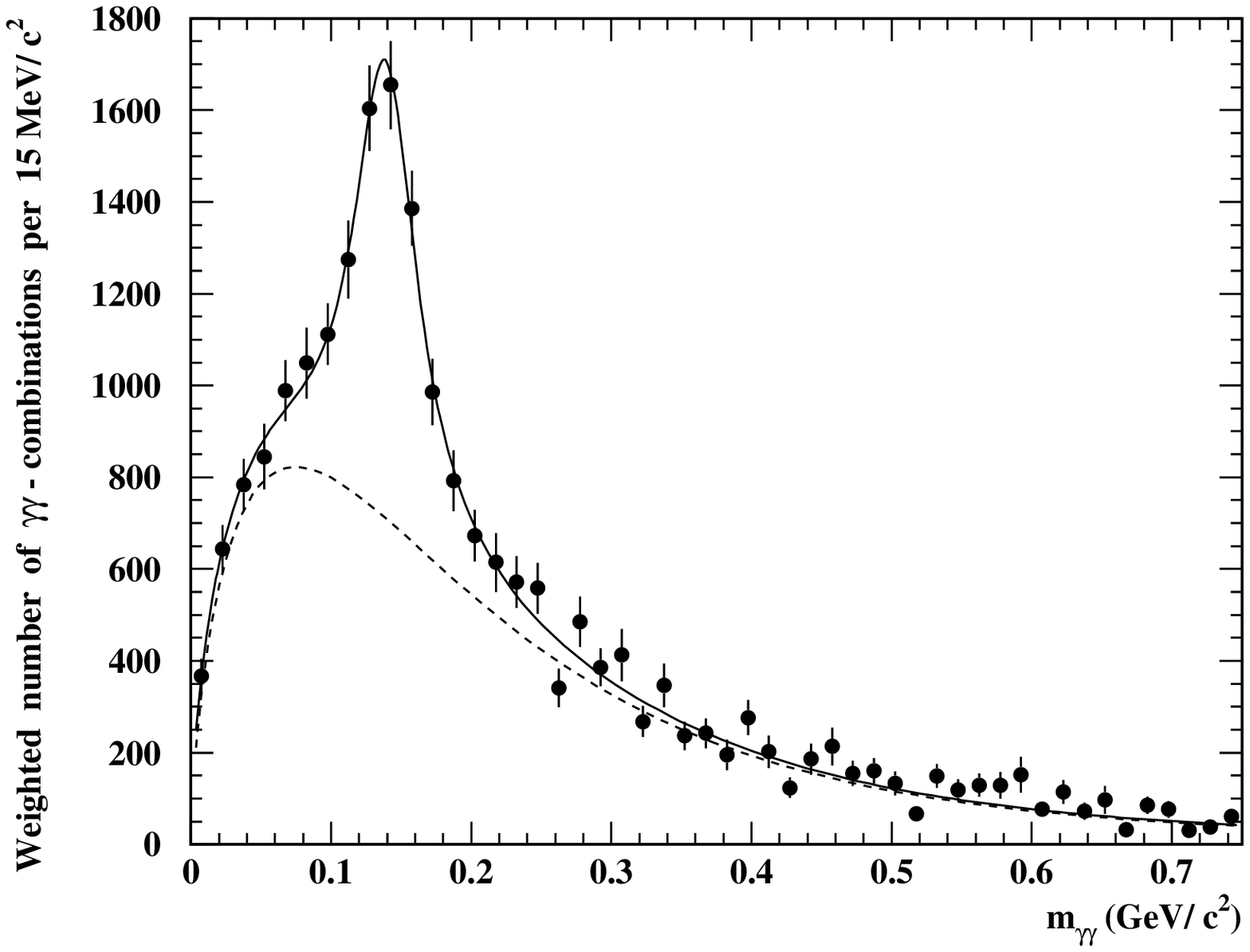}} \caption{The two-gamma effective mass
distribution. Solid curve: the fit result for the sum of
Breit-Wigner and background distributions; dashed curve: the
background contribution (see the text).}
\end{figure}

\newpage
\begin{figure}[ht]
\resizebox{0.9 \textwidth}{!}{\includegraphics*[bb=20 40 500 610]
{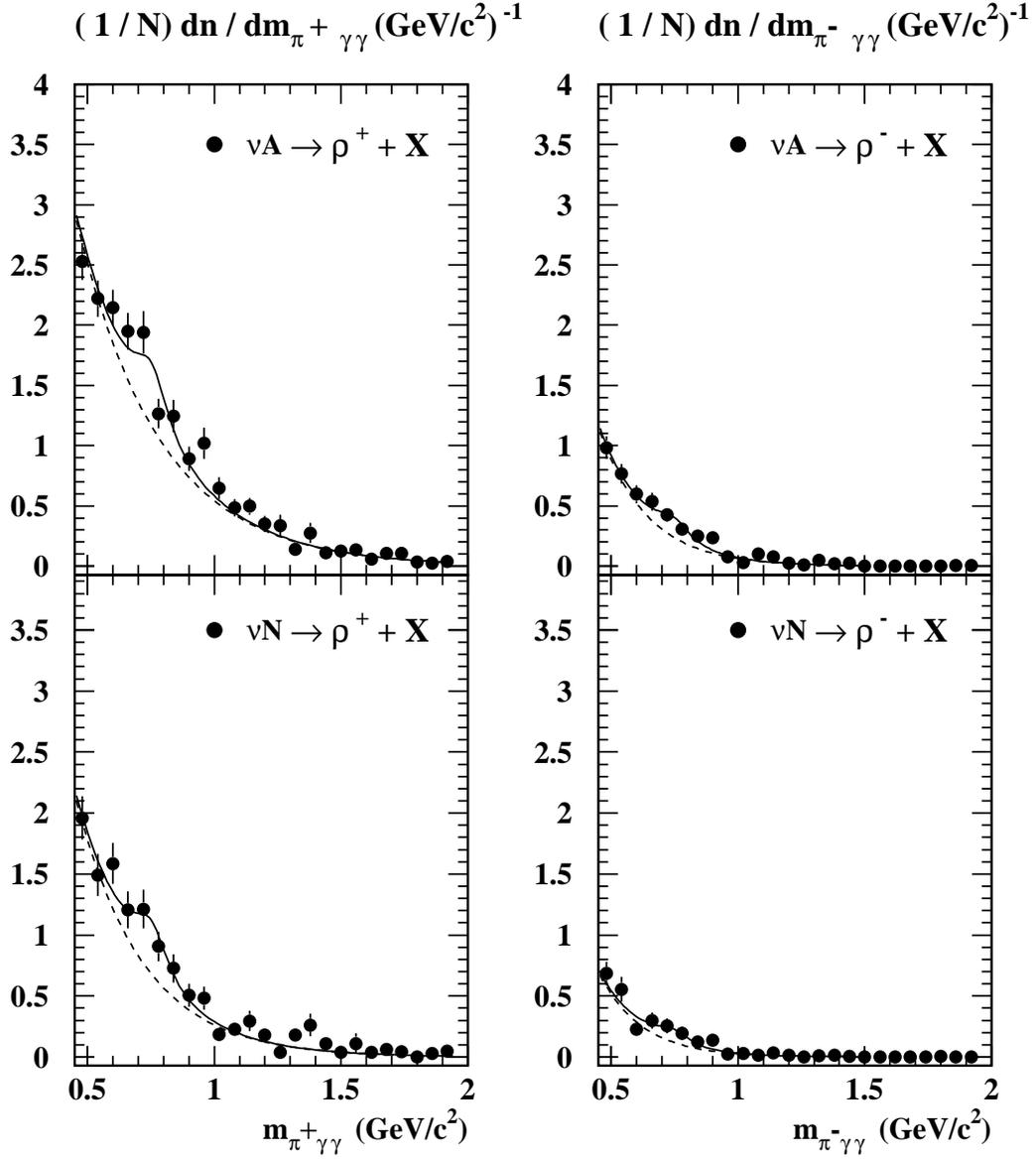}} \caption{The effective mass distribution for systems
($\pi^+ {\gamma\gamma}$) and ($\pi^-{\gamma\gamma}$) for nuclear
and quasinucleon subsamples. Solid curves: the result of fit by
expression (2); dashed curves: the background distribution (see
the text).}
\end{figure}

\newpage
\begin{figure}[h]
\resizebox{1.0 \textwidth}{!}{\includegraphics*[bb=50 55 600
570]{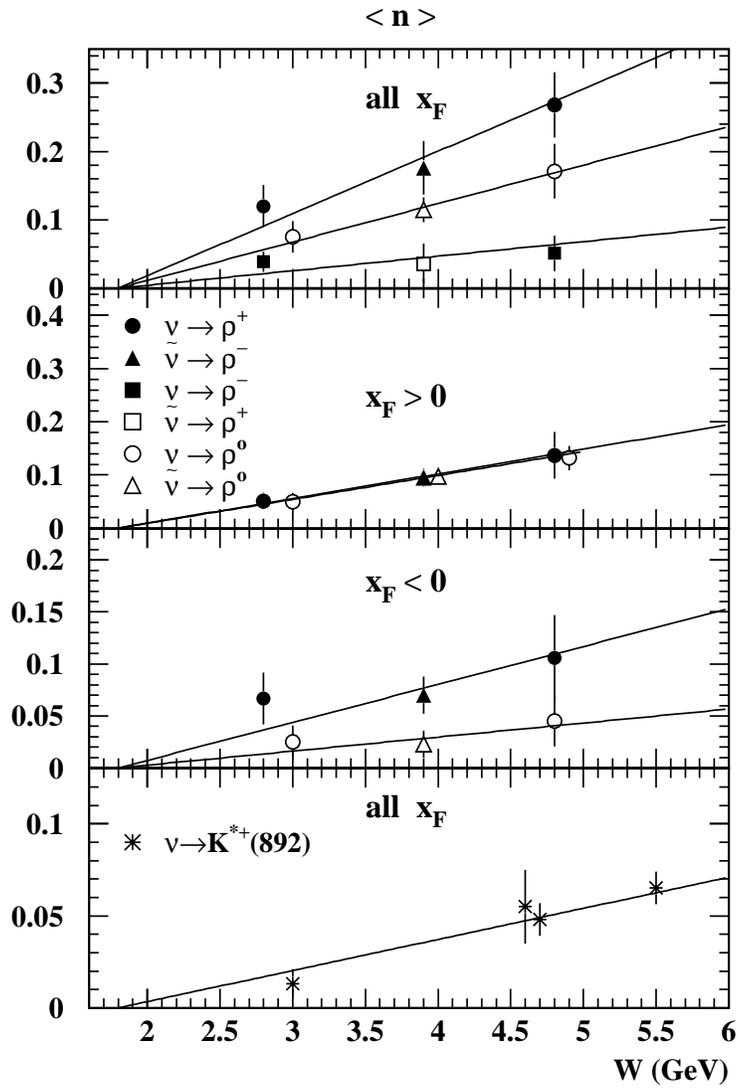}} \caption{The $W$ - dependence of $\rho$ and
$K^{*+}$(892) yields. Lines: the fit results (see the text).}
\end{figure}

\newpage
\begin{figure}[ht]
\resizebox{1.1 \textwidth}{!}{\includegraphics*[bb=10 150 600
570]{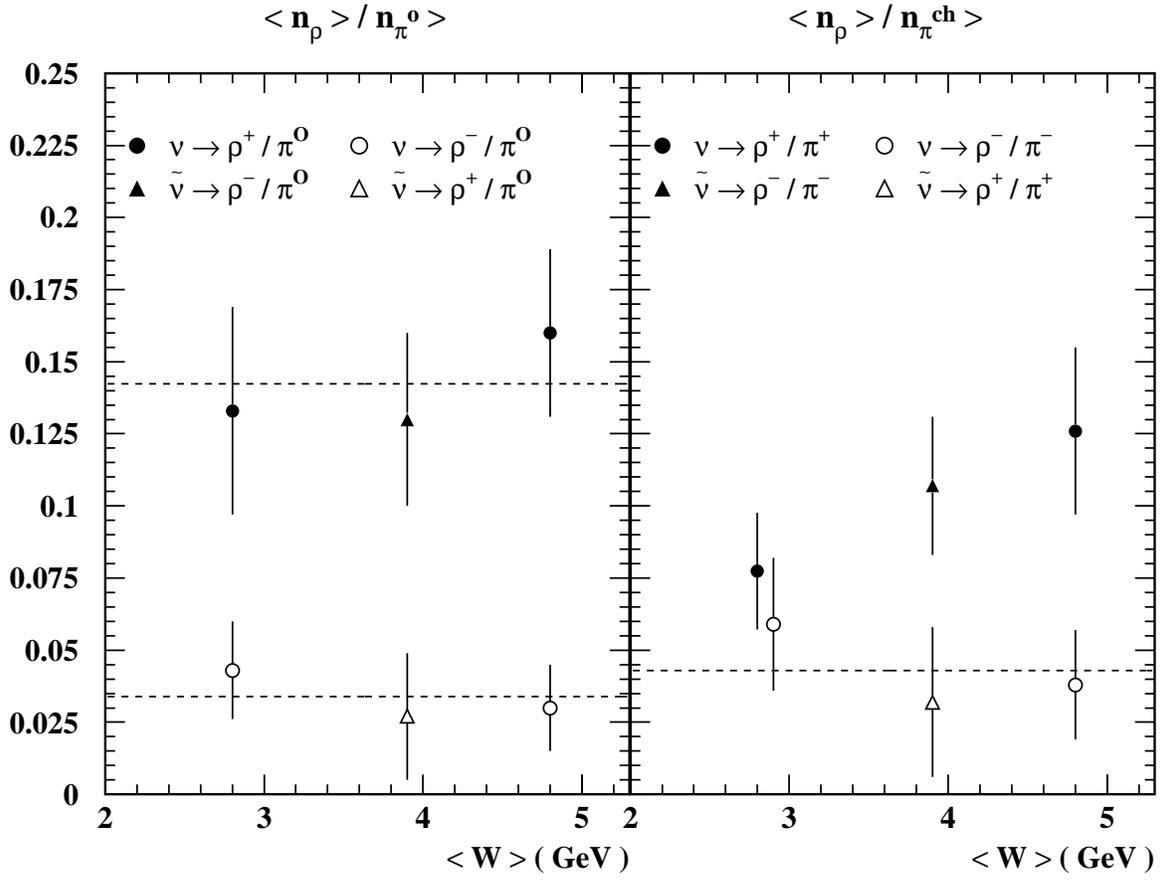}} \caption{The $W$ - dependence of the $\langle
n_{\rho} \rangle_A/\langle n_{\pi} \rangle_A$ ratios. Dashed
horizontal lines denote the mean values (averaged over the
considered $\langle W \rangle$ - range).}
\end{figure}
\vspace{0.5cm}

\newpage
\begin{figure}[ht]
\resizebox{0.9 \textwidth}{!}{\includegraphics*[bb=50 45 600
600]{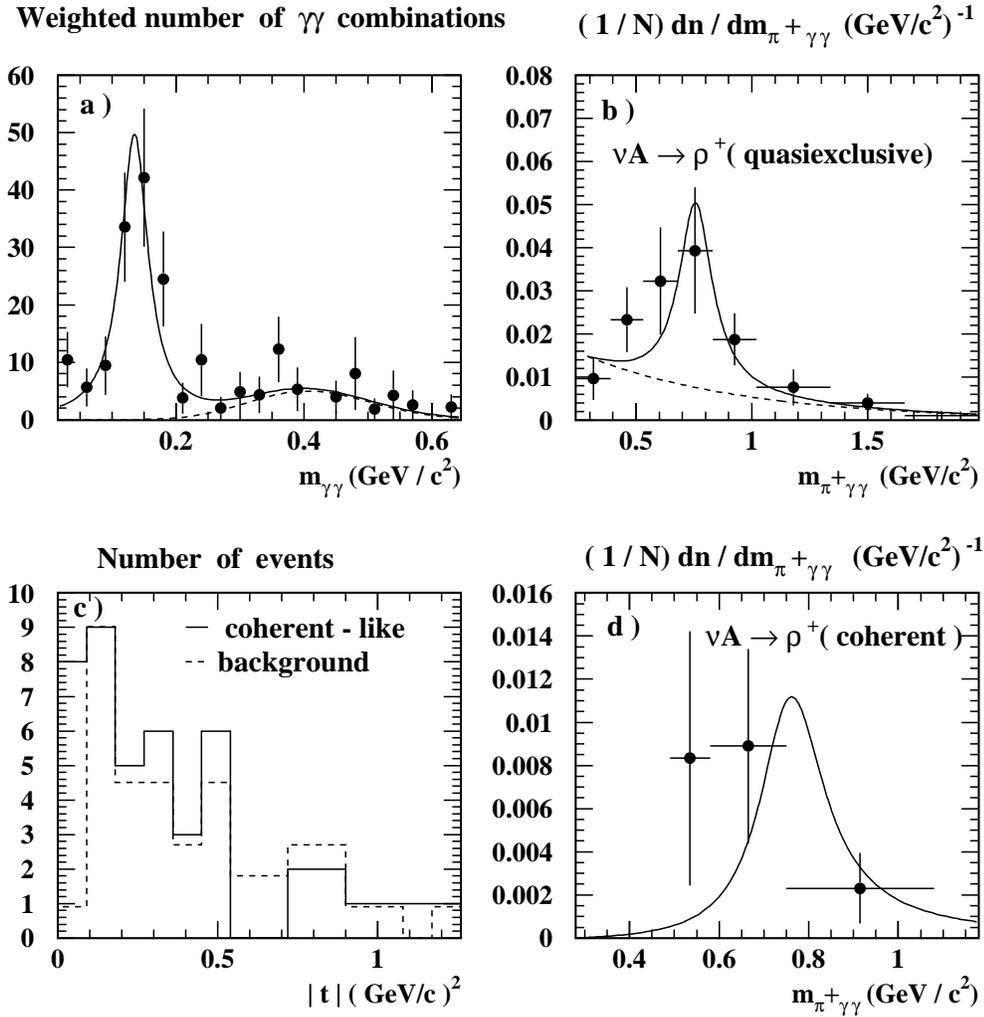}} \caption{a) two-gamma effective mass distribution
for reaction (4); the curves - as in Figure 1, but at fixed $m_0$
= 0.135 GeV$/c^2$ and $\Gamma_0$ = 0.055 GeV b) the effective mass
distribution for the system ($\pi^+{\gamma\gamma}$) in  the
reaction (4). The curves - as in Figure 2; c) the $|\,t|$ -
distributions for reaction (4) (solid histogram) and for
background events (dashed histogram, see the text); c) the
($\pi^+{\gamma\gamma}$) effective mass distribution for
coherent-like events. The curve is the fit result (see the text).}
\end{figure}

\vspace{0.5cm}

\newpage
\begin{figure}[ht]
\resizebox{0.9 \textwidth}{!}{\includegraphics*[bb=50 155 600
600]{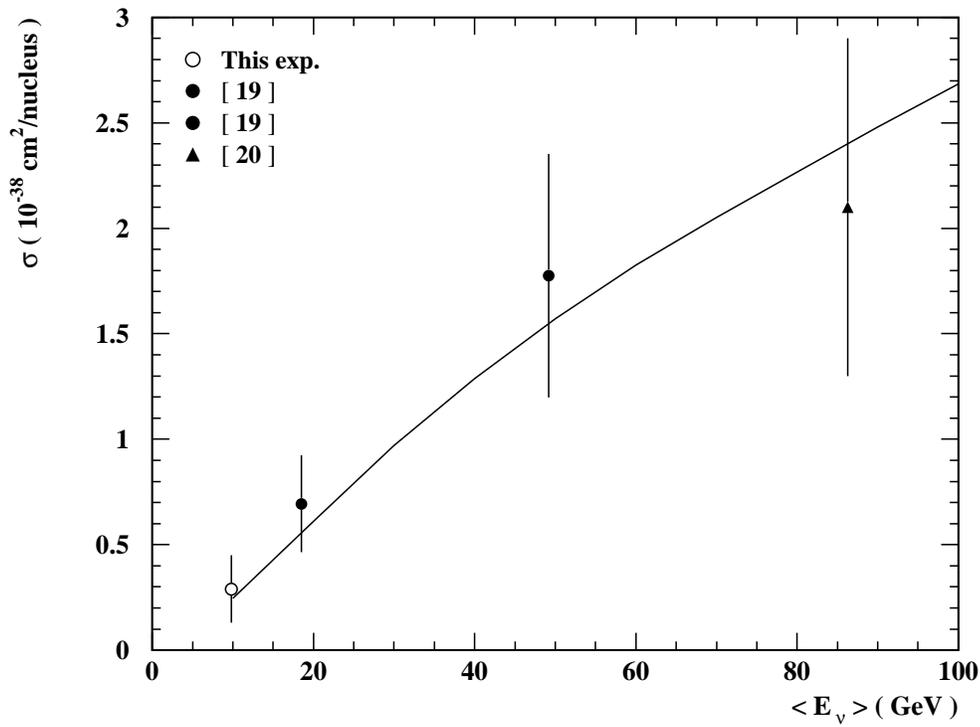}} \caption{The $\langle E_\nu \rangle$ - dependence
of the ratio $\sigma_{\rho}^{coh}$. The curve: the theoretical
prediction (taken from \cite{ref19}).}
\end{figure}

\end{document}